\documentclass{article}
\usepackage[utf8]{inputenc}

\usepackage{amsmath,amssymb,amsthm}
\usepackage{comment}
\usepackage{fullpage}
\usepackage{hyperref}
\usepackage{bbm}
\usepackage{color,graphicx,caption,subcaption}

\newtheorem*{proposition*}{Proposition}

\title{Discussion of ``Unbiased Markov chain Monte Carlo with couplings'' by Pierre E. Jacob, John O’Leary and Yves F. Atchad\'{e}}
\author{Dr Leah F. South$^1$\thanks{\href{mailto:l.south@lancaster.ac.uk}{l.south@lancaster.ac.uk}}, Dr Chris Nemeth$^1$, Professor Chris. J. Oates$^{2,3}$ \\ \small
$^1$Lancaster University, UK \\ \small
$^2$Newcastle University, UK  \\ \small
$^3$The Alan Turing Institute, UK}
\date{January 2020}

\begin{document}

\maketitle

The authors should be congratulated on an impressive and thought-provoking contribution to the field.
Traditional MCMC has benefited from the development of gradient-based control variates \cite{Assaraf,Barp,Mira,Oates,South}, but it may be more difficult to design gradient-based control variates for Unbiased MCMC.
Following the notation in the paper, let $\hat{\pi}^R(h) := \frac{1}{R} \sum_{r=1}^R H_{k:m}^{(r)}(X,Y)$ and $\pi(h) := \mathbb{E}[h(X)]$.
Under Assumptions 1-3, Proposition 1 establishes that $\sigma(h)^2 := \mathbb{V}[H_{k:m}(X,Y)] < \infty$, so that
$$
\sqrt{R} (\hat{\pi}^R(h) - \pi(h)) \stackrel{d}{\rightarrow} N(0,\sigma(h)^2)
$$ 
as $R \rightarrow \infty$.
A control variate $g$ should therefore be selected such that $\pi(g) = 0$ and $\sigma(h - g) \ll \sigma(h)$.
In (3.2) it was demonstrated that, in the large $m$ and $k$ limit, the quantity $\sigma(h)^2$ is just the asymptotic variance from traditional MCMC; existing gradient-based control variates can therefore be used \cite{Belomestny,Mijatovic}.
However, at finite $m$ and $k$ the dependence of $\sigma(h)$ on $h$ is far from explicit. 
One could use sample-splitting to construct an approximation of the form
$$
\hat{\sigma}(h)^2 = \frac{1}{\lfloor R / 2 \rfloor} \sum_{r = 1}^{\lfloor R / 2 \rfloor} \Bigg( H_{k:m}^{(r)}(X,Y) - \frac{1}{\lfloor R / 2 \rfloor} \sum_{r' = 1}^{\lfloor R / 2 \rfloor} H_{k:m}^{(r')}(X,Y) \Bigg)^2
$$
and attempt to minimise $\hat{\sigma}(h-g)$. 
Alternatively, one could bound $\sigma(h-g)$ in terms of quantities that are independent of the Markov chain and then minimise the bound.
One such bound is provided in the following result, stated for $k = m = 0$ for simplicity, which we do not claim to be in any sense optimal:

Let Assumptions 1-3 be satisfied, with $\eta$ as in Assumption 1 and $C$, $\delta$ as in Assumption 2.
Let $\pi_t$ be the law of $X_t$ and assume that $\lambda := \sup_{t \geq 0} d_{\textsc{TV}}(\pi,\pi_t) < \infty$.
Let $\mathcal{H}$ be a reproducing kernel Hilbert space, with norm denoted $\|\cdot\|_{\mathcal{H}}$ and with kernel $K : \mathcal{X} \times \mathcal{X} \rightarrow \mathbb{R}$ satisfying $K(x,x) \leq 1$ for all $x \in \mathcal{X}$.
If $|h|^{2 + \eta} \in \mathcal{H}$ then
\begin{equation}
\sigma(h) \leq \gamma \left( \pi(|h|^{2 + \eta}) + \lambda \||h|^{2 + \eta}\|_{\mathcal{H}} \right)^{\frac{1}{2 + \eta}} + \mathbb{E}[ h(X_0)^2 ]^{1/2}, \qquad \gamma^2 = 4 C^{\frac{\eta}{2+\eta}} \frac{\delta^{\eta / (2 + \eta)} }{ ( 1 - \delta^{\eta / (4 + 2\eta)} )^2 },
\label{eq: bound on sigma}
\end{equation}
where the positive constants $\gamma$ and $\lambda$ are $h$-independent.
The proof is provided in Appendix \ref{app:upper bound}.
None of the three $h$-dependent quantities in the bound depend on the law of the Markov chain and thus minimisation of the bound may be practical.
The, in practice unknown, values of $\gamma(\delta)$ and $\lambda$ determine which of the three terms dominate the bound.

\begin{figure}
\centering
\begin{subfigure}[b]{0.3\textwidth}
\includegraphics[width=1\textwidth]{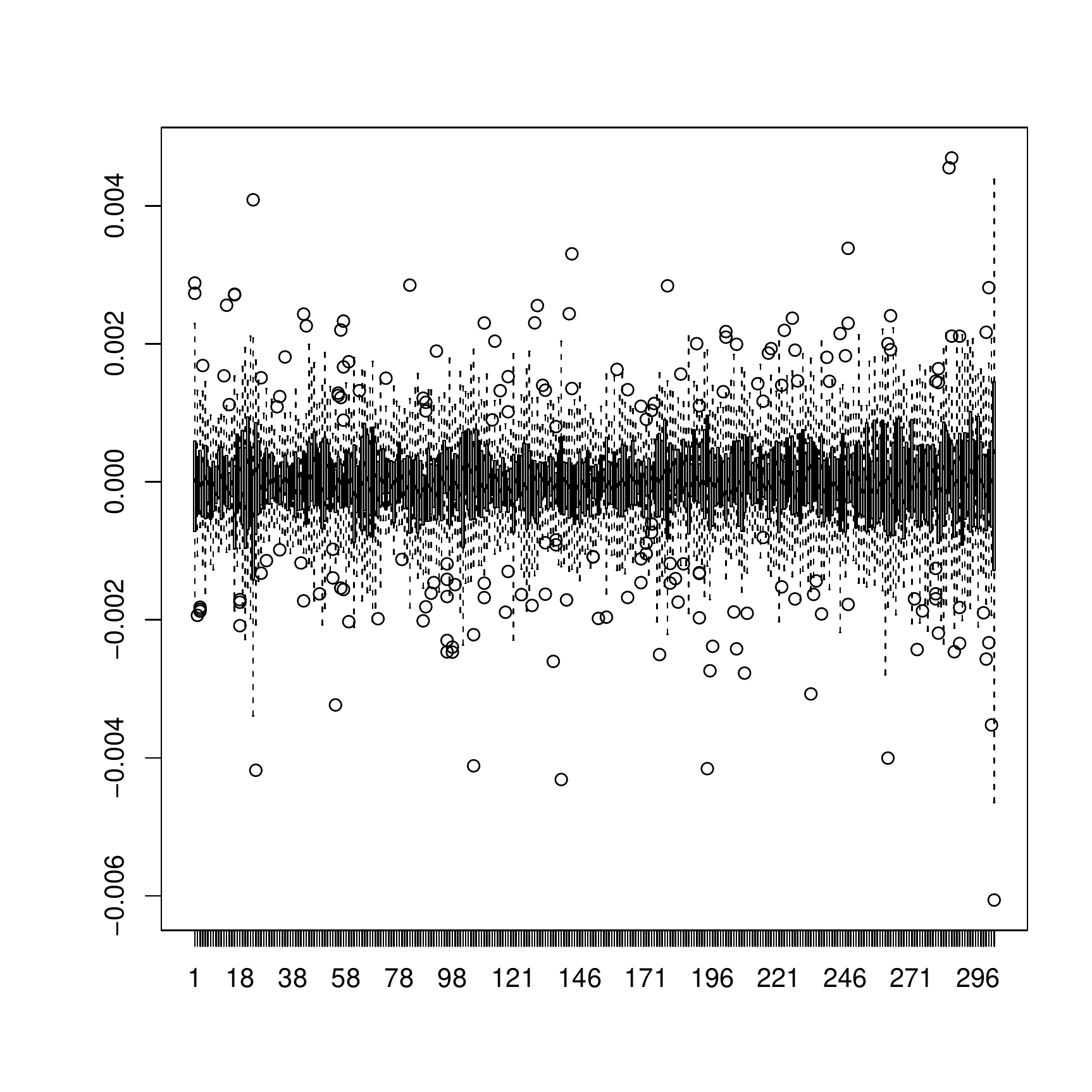}
\caption{Unbiased MCMC}
\end{subfigure}
\begin{subfigure}[b]{0.3\textwidth}
\includegraphics[width=1\textwidth]{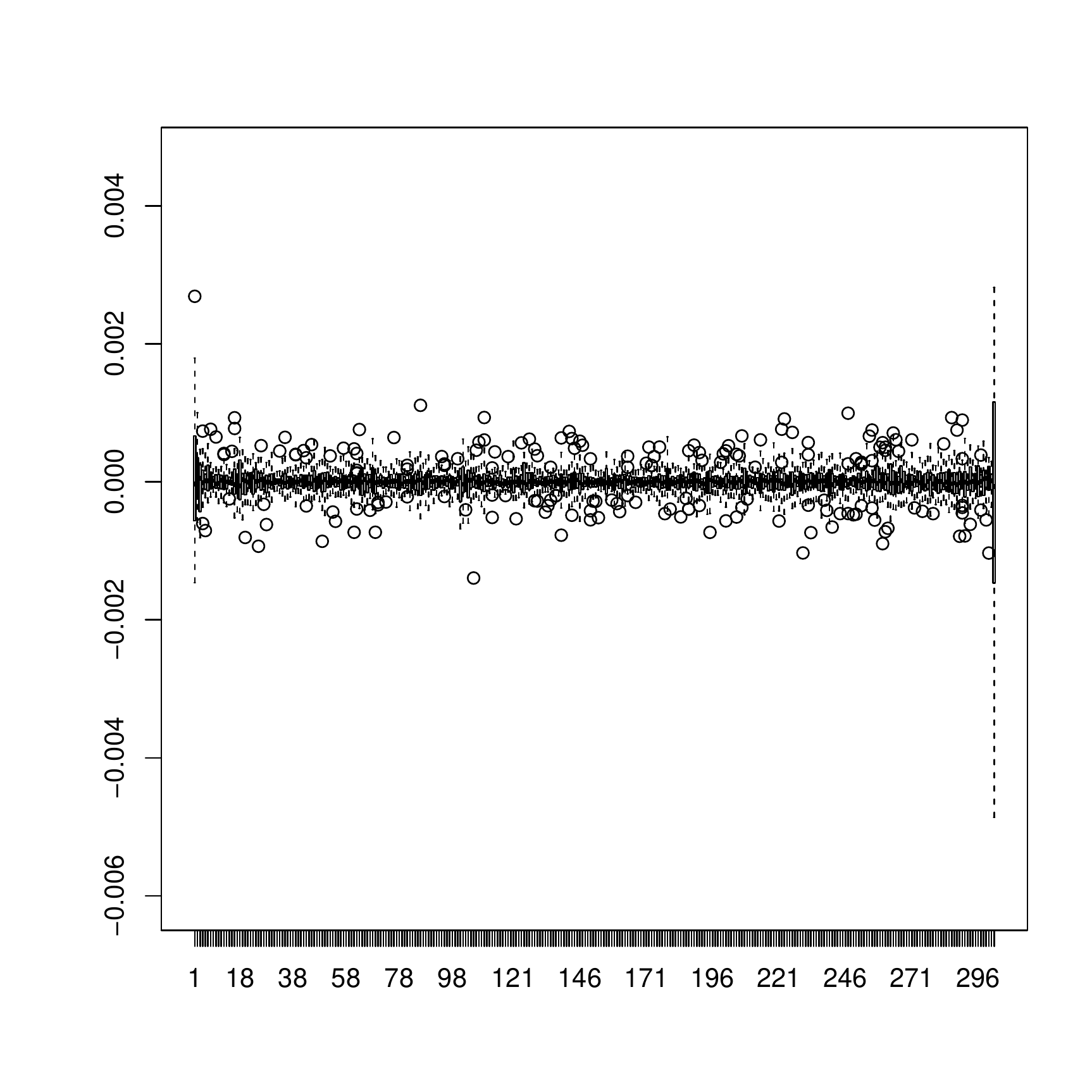}
\caption{Approach (i)}
\end{subfigure}
\begin{subfigure}[b]{0.3\textwidth}
\includegraphics[width=1\textwidth]{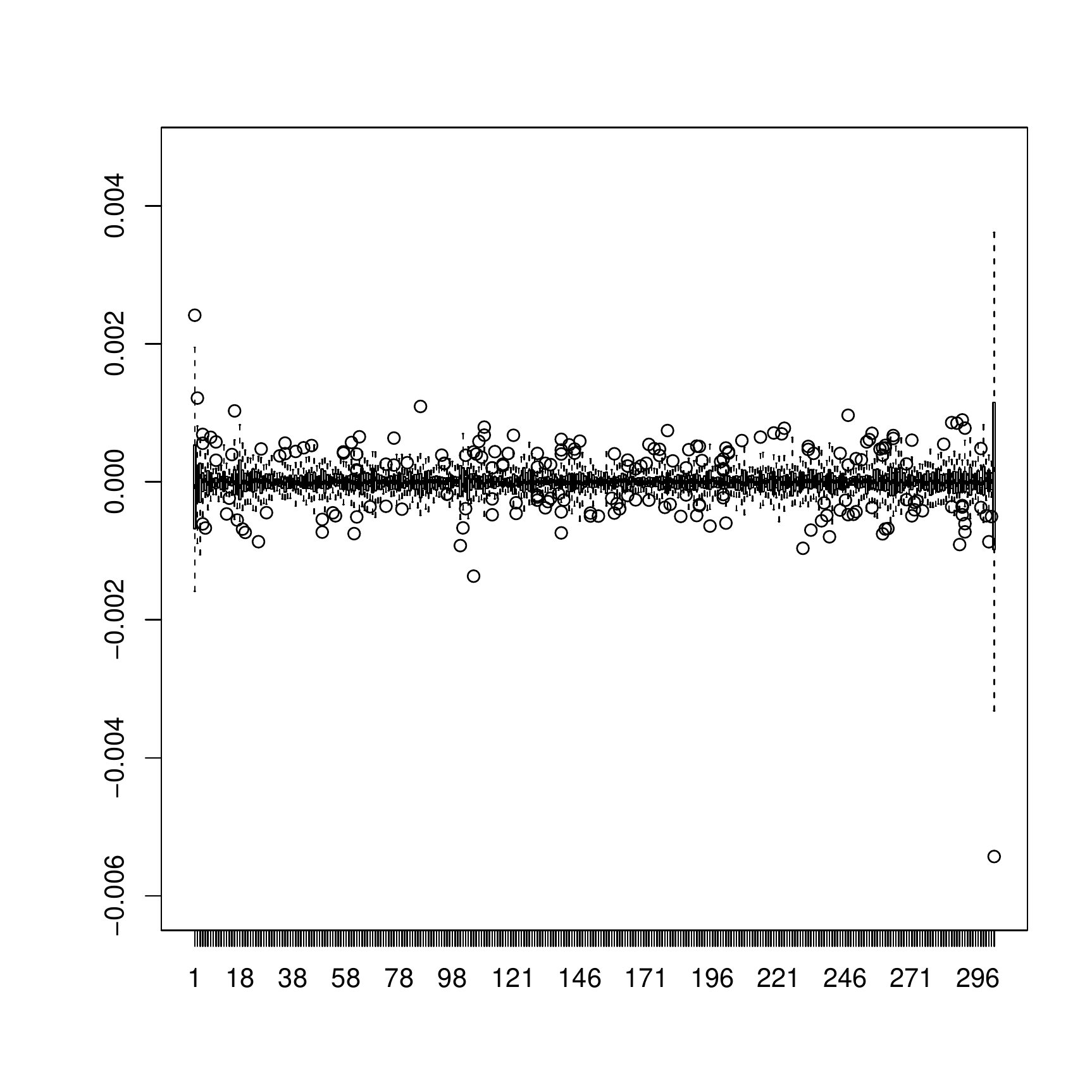}
\caption{Approach (ii)}
\end{subfigure}
\caption{Boxplots of 30 estimates of the marginal posterior expectations $\mathbb{E}[X_i]$ for $i=1,\ldots,302$ in the logistic regression example of \cite{Heng}. The following strategies were compared:
(i) direct minimisation of $\hat{\sigma}(h-g)$, and (ii) minimisation of the bound \eqref{eq: bound on sigma} on $\sigma(h-g)$, with $\eta, \lambda \ll 1$, $\gamma \gg 1$ and $\pi$ approximated with MCMC output. In each case $g$ was a first order Stein control variate \cite{Mira,South} estimated using the first $\lfloor R / 2 \rfloor$ chains, whilst $\pi(h)$ was estimated using the remaining $R - \lfloor R / 2 \rfloor$ chains in order that the estimators remain unbiased. Runs are based on $k=330$, $m=3300$ and $R=32$. The empirical means from approach (ii) are subtracted for visualisation purposes.
The median variance reduction factor using approaches (i) and (ii) is approximately 20.
However, approach (i) depended strongly on the numerical approach used to minimise the non-convex objective function $\hat{\sigma}( h-g)^2$. Code to reproduce the experiment is provided at \url{https://github.com/LeahPrice/debiasedhmc}.
}
\label{fig: CV results}
\end{figure}

Figure \ref{fig: CV results} displays the variance reduction achieved in the 302-dimensional logistic regression example of \cite{Heng}. 
These results are encouraging, but more work is required to develop an understanding of gradient-based control variates for Unbiased MCMC.

\newpage

\appendix
\section{Derivation of the Upper Bound}\label{app:upper bound}

The aim in what follows is to reproduce the proof of Proposition 1 in \cite{Jacob} whilst explicitly tracking the terms that are $h$-dependent.
To avoid reproducing large amounts of \cite{Jacob}, we assume familiarity with the notation and quantities defined in that work.

The first part of the argument in \cite{Jacob} uses Assumption 1 to deduce that $\mathbb{E}[\Delta_t^2] \leq \tilde{C} \tilde{\delta}^t$ for some $\tilde{C}$ and all $t \geq 0$. 
Our first task is to explicitly compute the constant $\tilde{C}$ in terms of the quantities $\eta$ and $D$ in Assumption 1.
To this end, we reproduce the argument alluded to in the paper:
\begin{eqnarray*}
( \mathbb{E}[ |\Delta_t|^{2 + \eta} ] )^{\frac{1}{2 + \eta}} & = & ( \mathbb{E}[ | h(X_t) - h(Y_{t-1}) |^{2 + \eta} ] )^{\frac{1}{2 + \eta}} \\
& \leq & ( \mathbb{E}[ | h(X_t) |^{2 + \eta} ] )^{\frac{1}{2 + \eta}} + ( \mathbb{E}[ | h(Y_{t-1}) |^{2 + \eta} ] )^{\frac{1}{2 + \eta}} \qquad \text{(Minkowski's inequality)} \\
& \leq & D^{\frac{1}{2 + \eta}} + D^{\frac{1}{2 + \eta}} \qquad \text{(Assumption 1)} \\
\implies \qquad
\mathbb{E}[\Delta_t^2] \; = \; \mathbb{E}[\Delta_t^2 1(\tau > t) ] & \leq & \mathbb{E}[ |\Delta_t|^{2 + \eta} ]^{\frac{2}{2 + \eta}} \mathbb{E}[1(\tau > t)]^{\frac{\eta}{2 + \eta}} \qquad \text{(H\"{o}lder's inequality)} \\
& \leq & \big( 2D^{\frac{1}{2 + \eta}} \big)^2 (C \delta^t)^{\frac{\eta}{2+\eta}} \qquad \text{(Assumption 2)} \\
& = & 4 C^{\frac{\eta}{2+\eta}} D^{\frac{2}{2+\eta}} \tilde{\delta}^t \; = \; \tilde{C} \tilde{\delta}^t, \qquad \tilde{C} = 4 C^{\frac{\eta}{2+\eta}} D^{\frac{2}{2+\eta}} .
\end{eqnarray*}

It is then stated in the proof of Proposition 1 in \cite{Jacob} that $\mathbb{E}[ ( H_0^{n'}(X,Y) - H_0^n(X,Y) )^2 ] \leq \bar{C} \tilde{\delta}^n$ where $\tilde{\delta} \in (0,1)$ for some $\bar{C}$ and all $n,n'$ with $n' \geq n$; we reproduce the implied argument to explicitly represent $\bar{C}$ in terms of $\eta$ and $D$ next:
\begin{eqnarray*}
\mathbb{E}[ ( H_0^{n'}(X,Y) - H_0^n(X,Y) )^2 ] & = & \sum_{s=n+1}^{n'} \sum_{t=n+1}^{n'} \mathbb{E}[\Delta_s \Delta_t] \\
& \leq & \sum_{s=n+1}^{n'} \sum_{t=n+1}^{n'} \mathbb{E}[\Delta_s^2]^{1/2} \mathbb{E}[\Delta_t^2]^{1/2} \qquad \text{(Cauchy-Schwarz inequality)} \\
& \leq & \sum_{s=n+1}^{n'} \sum_{t=n+1}^{n'} (\tilde{C} \tilde{\delta}^s)^{1/2} (\tilde{C} \tilde{\delta}^t)^{1/2} \\
& = & \tilde{C} \sum_{s=n+1}^{n'} (\tilde{\delta}^{1/2})^{s+n+1} \sum_{t=0}^{n'-n-1} (\tilde{\delta}^{1/2})^t \\
& = & \tilde{C} \sum_{s=n+1}^{n'} (\tilde{\delta}^{1/2})^{s+n+1} \left( \frac{1 - (\tilde{\delta}^{1/2})^{n'-n} }{ 1 - \tilde{\delta}^{1/2} } \right) \\
& \leq & \tilde{C} \frac{1}{ 1 - \tilde{\delta}^{1/2} } \sum_{s=n+1}^{n'} (\tilde{\delta}^{1/2})^{s+n+1}  \\
& = & \tilde{C} \frac{1}{ 1 - \tilde{\delta}^{1/2} } (\tilde{\delta}^{1/2})^{2n+2} \sum_{s=0}^{n'-n-1} (\tilde{\delta}^{1/2})^s \\
& = & \tilde{C} \frac{1}{ 1 - \tilde{\delta}^{1/2} } (\tilde{\delta}^{1/2})^{2n+2} \left( \frac{ 1 - (\tilde{\delta}^{1/2})^{n'-n} }{ 1 - \tilde{\delta}^{1/2} } \right) \; \leq \; \tilde{C} \frac{\tilde{\delta}}{ ( 1 - \tilde{\delta}^{1/2} )^2 } \tilde{\delta}^n
\end{eqnarray*}
so we may take 
\begin{eqnarray}
\bar{C} = \frac{\tilde{\delta}}{(1 - \tilde{\delta}^{1/2})^2} \tilde{C} = \frac{\tilde{\delta}}{(1 - \tilde{\delta}^{1/2})^2} \times 4 C^{\frac{\eta}{2+\eta}} D^{\frac{2}{2 + \eta}} = \gamma^2 D^{\frac{2}{2 + \eta}}, \qquad \gamma^2 := 4 C^{\frac{\eta}{2+\eta}} \frac{\delta^{\frac{\eta}{2+\eta} }}{( 1 - \delta^{\frac{\eta}{4+2\eta}} )^2} \label{bound on barC}
\end{eqnarray}
where $\gamma$ is a $h$-independent constant that depends only on the law of the meeting time for the Markov chains. 
The constant $\gamma$ is finite since $\tilde{\delta} \in (0,1)$.

The stylised bound that we present is rooted in the concept of the \emph{maximum mean discrepancy} $d_{\mathcal{H}}$ associated to the reproducing kernel Hilbert space $\mathcal{H}$, defined as
$$
d_{\mathcal{H}}(\pi,\pi') := \sup_{\|f\|_{\mathcal{H}} \leq 1} |\pi(f) - \pi'(f)| .
$$
If $|h|^{2 + \eta} \in \mathcal{H}$ then we have from the definition of the maximum mean discrepancy that
$$
|\pi(|h|^{2 + \eta}) - \pi'(|h|^{2 + \eta})| \leq \||h|^{2 + \eta}\|_{\mathcal{H}} d_{\mathcal{H}}(\pi,\pi').
$$
Taking $\pi' = \pi_t$ to be the law of $X_t$ thus gives that
\begin{eqnarray*}
|\pi(|h|^{2 + \eta}) - \mathbb{E}[|h(X_t)|^{2 + \eta})] & \leq & \||h|^{2 + \eta}\|_{\mathcal{H}} d_{\mathcal{H}}(\pi,\pi_t) \\
\implies \qquad \mathbb{E}[|h(X_t)|^{2 + \eta}] & \leq & \pi(|h|^{2 + \eta}) + \||h|^{2 + \eta}\|_{\mathcal{H}} d_{\mathcal{H}}(\pi,\pi_t) \\
\implies \qquad \sup_{t \geq 0} \mathbb{E}[|h(X_t)|^{2 + \eta}] & \leq & \pi(|h|^{2 + \eta}) + \||h|^{2 + \eta}\|_{\mathcal{H}} \sup_{t \geq 0} d_{\mathcal{H}}(\pi,\pi_t) .
\end{eqnarray*}
Thus we may take the constant $D$ in Assumption 1 to be
\begin{eqnarray}
D = \pi(|h|^{2 + \eta}) + \||h|^{2 + \eta}\|_{\mathcal{H}} \sup_{t \geq 0} d_{\mathcal{H}}(\pi,\pi_t) . \label{D bound}
\end{eqnarray}
In what follows we let $\lambda_{\mathcal{H}} := \sup_{t \geq 0} d_{\mathcal{H}}(\pi,\pi_t)$ be a $h$-independent constant that depends on the law of the Markov chain used.
It is necessary to check that $\lambda_{\mathcal{H}}$ is finite.
Let $\langle \cdot , \cdot \rangle_{\mathcal{H}}$ be the inner product in $\mathcal{H}$.
The assumption that $\mathcal{H}$ is a reproducing kernel Hilbert space means that $|h(x)| = |\langle h , K(\cdot,x) \rangle_{\mathcal{H}}| \leq \|h\|_{\mathcal{H}} \|K(\cdot,x)\|_{\mathcal{H}} = K(x,x)^{1/2} \|h\|_{\mathcal{H}}$ , from the reproducing property and Cauchy-Schwarz.
Since the kernel $K$ was assumed to satisfy $K(x,x) \leq 1$, it follows that $\|h\|_\infty \leq \|h\|_{\mathcal{H}}$.
Thus 
\begin{eqnarray*}
d_{\mathcal{H}}(\pi,\pi') = \sup_{\|h\|_{\mathcal{H}} \leq 1} | \pi(h) - \pi'(h) | \leq \sup_{ \|h\|_\infty \leq 1} | \pi(h) - \pi'(h) | = d_{\text{TV}}(\pi,\pi') .
\end{eqnarray*}
Thus $\lambda_{\mathcal{H}} \leq \lambda < \infty$ as required.

To complete the argument we proceed as follows:
\begin{eqnarray*}
\mathbb{E}[ ( H_0^{n'}(X,Y) - H_0^n(X,Y) )^2 ] & \leq & \bar{C} \tilde{\delta}^n \\
\implies \qquad | \mathbb{E}[ H_0^{n'}(X,Y)^2 ]^{1/2} - \mathbb{E}[ (H_0^n(X,Y))^2 ]^{1/2} | & \leq & ( \bar{C} \tilde{\delta}^n )^{1/2} \qquad \text{(reverse Minkowski inequality)} \\
\implies \qquad \mathbb{E}[ H_0(X,Y)^2 ]^{1/2} & \leq & \bar{C}^{1/2} + \mathbb{E}[ h(X_0)^2 ]^{1/2} \qquad \text{(taking $n=0$, $n'=\infty$)} \\
\implies \qquad \sigma(h) & \leq & \bar{C}^{1/2} + \mathbb{E}[ h(X_0)^2 ]^{1/2} \qquad \text{(since $\mathbb{V}[Z] \leq \mathbb{E}[Z^2]$)} \\
& \leq & \gamma D^{\frac{1}{2 + \eta}} + \mathbb{E}[ h(X_0)^2 ]^{1/2} \qquad \text{(from \eqref{bound on barC})} \\
& \leq & \gamma \left( \pi(|h|^{2 + \eta}) + \lambda \||h|^{2 + \eta}\|_{\mathcal{H}} \right)^{\frac{1}{2 + \eta}} + \mathbb{E}[ h(X_0)^2 ]^{1/2}
\end{eqnarray*}
where the final line follows from \eqref{D bound} and the fact that $\lambda_{\mathcal{H}} \leq \lambda$.

\end{document}